\documentclass[sigconf]{acmart}
\AtBeginDocument{%
  }
\acmConference[CHI '26 Workshop on Data Literacy]{CHI 2026 Workshop on Data Literacy}{April 13, 2026}{Barcelona, Spain}
\setcopyright{none}
\acmISBN{}
\acmDOI{}
\acmPrice{}
\renewcommand\footnotetextcopyrightpermission[1]{}

\begin{document}
\makeatletter
\def\@acmBooktitle{CHI 2026 Workshop on Data Literacy, April 13, 2026, Barcelona, Spain}
\makeatother

\title{Visualization Retrieval for Data Literacy: Position Paper}

\author{Huyen N. Nguyen}
\email{huyen\_nguyen@hms.harvard.edu}
\orcid{0000-0001-6554-2327}
\affiliation{%
  \institution{Harvard Medical School}
  \city{Boston}
  \state{Massachusetts}
  \country{USA}
}

\author{Nils Gehlenborg}
\email{nils@hms.harvard.edu}
\orcid{0000-0003-0327-8297}
\affiliation{%
  \institution{Harvard Medical School}
  \city{Boston}
  \state{Massachusetts}
  \country{USA}
}

\renewcommand{\shortauthors}{Nguyen and Gehlenborg}

\begin{abstract}

Current resources for data literacy education, such as visualization galleries and datasets, provide useful examples but lack mechanisms for learners to query, compare, and navigate the visualization design space efficiently. This position paper advocates for visualization retrieval as essential infrastructure for data literacy, transforming static collections into dynamic, inquiry-based learning environments. We analyze the role of retrieval across the data lifecycle, demonstrating how it facilitates design space exploration and vocabulary expansion, supports data consumption through visualization comparison and critique, and aids data management via resource curation. We outline key opportunities for future research and system design, including integrated retrieval-authoring environments, pedagogical relevance modeling, and collaborative educational corpora. Ultimately, we argue that visualization retrieval systems empower learners to articulate intent, bridge technical barriers, and proactively reason with data.

\end{abstract}



\maketitle

\section{Introduction}

Recent research in data science education emphasizes the importance of real-world examples for developing data literacy. Within this space, data visualization acts as both a motivational tool~\cite{majumder2025developing} and a means of communication~\cite{roberts2025data}. The concept of \textit{visualization literacy} has emerged as a distinct competency, encompassing the ability to read, interpret, and produce visual representations of data~\cite{borner2019data}. Although growing collections of visualization databases and datasets~\cite{Koenen_DaVE, Hutchinson_dataset, VisImages, Beagle, walters2025gqvis, chen2021vis30k} provide valuable pedagogical resources, these repositories remain largely static and curator-driven. Most importantly, the ability to retrieve specific, relevant examples is currently limited.

This lack of retrieval capability prevents students from navigating the design space more effectively. Prior work shows that exposure to relevant visualization examples helps participants understand which visual encodings and interactions are possible for their specific problems~\cite{nguyen_geranium}, and that visualization designers routinely search for visual examples as sources of inspiration~\cite{baigelenov2025visualization}. To facilitate this, learners need mechanisms to ask questions of these collections, using either images or text keywords to drive their exploration. Aligning with the principles of inquiry-based learning~\cite{lazonder2016meta}, visualization retrieval offers a promising pathway for data literacy education. However, current systems lack the necessary support for exploring datasets through visual evidence, and ultimately navigating design spaces intentionally and proactively. 

In this position paper, we argue that \textit{visualization retrieval} is an essential, yet underexplored, infrastructure for data literacy. We address this gap by (1) contextualizing visualization retrieval within data literacy education, (2) mapping its role across the data lifecycle---production, consumption, and management, and (3) outlining opportunities for system design and research that integrate retrieval into teaching practice.

\section{Background: Visualization Retrieval}

\begin{figure*}[!h] 
  \centering
  \includegraphics[width=.88\textwidth]{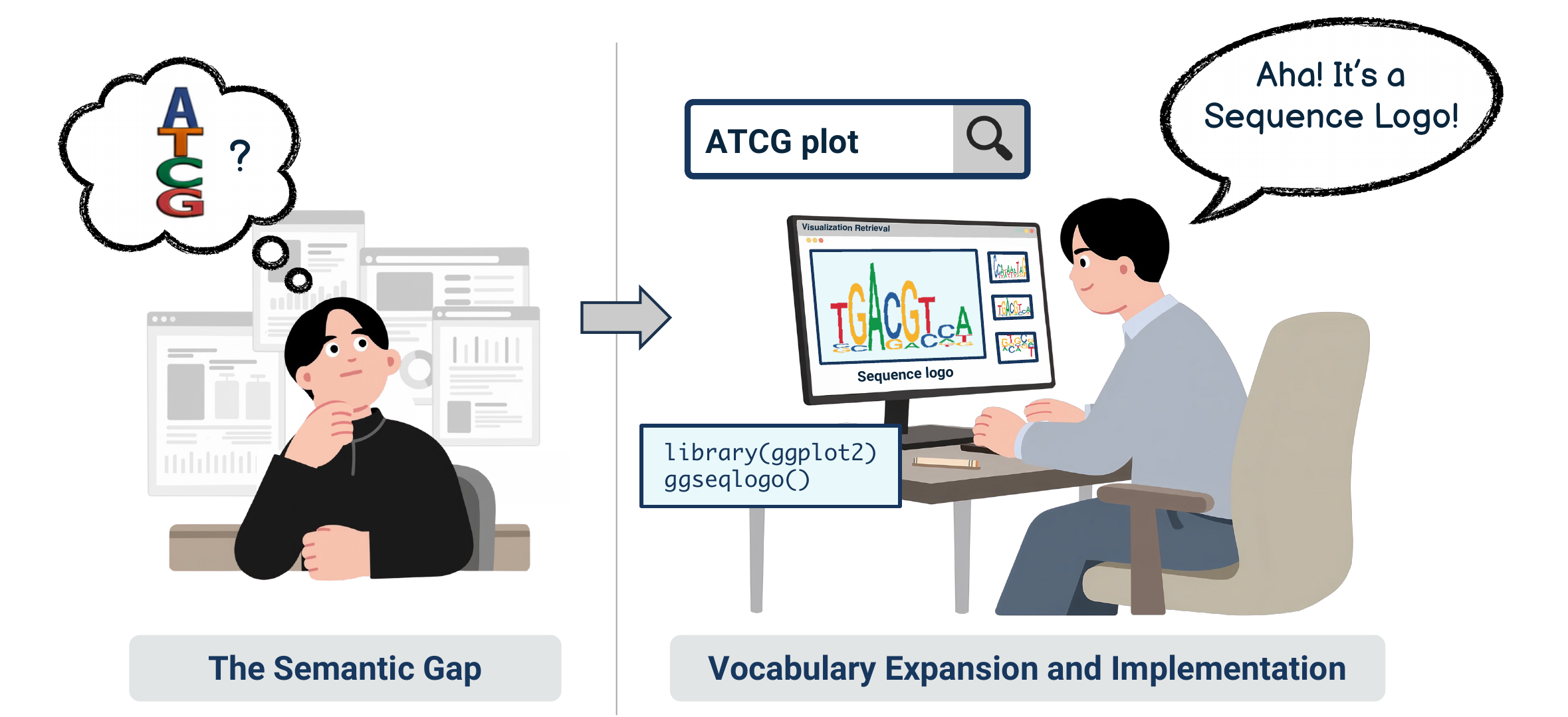}
  \caption{Illustration of visualization retrieval bridging the semantic gap. 
A student searching for an \textit{``ATCG plot''} lacks the formal vocabulary to find \textit{``sequence logo''} visualizations. A visualization retrieval system can match this informal query to sequence logo examples and return them with the correct terminology and associated code, helping the learner recognize the appropriate chart type, expand their visual vocabulary, and see how such a visualization can be implemented.}
\Description{Two-panel figure. Left panel shows a learner imagining stacked DNA letters labeled 'A T C G' while browsing a static gallery of unrelated charts. Right panel shows the learner entering 'A T C G plot' into a visualization retrieval interface and receiving sequence logo examples with associated code and the label 'sequence logo.'}
  \label{fig:example}
\end{figure*}

Information retrieval systems are the infrastructure behind modern knowledge acquisition, yet the application of these principles to data visualization remains a specialized challenge. As defined in foundational information retrieval literature~\cite{manning2008introduction}, the core objective is to find relevant documents that satisfy a user's information need through query mechanisms. In the context of visualization, this involves addressing the semantic gap: the challenge of translating user intent (e.g., \textit{``I want to show a correlation''}) into low-level visual features (e.g., scatterplots, regression lines) that a system can index. For example, a novice student in genomics might type \textit{``A T C G plot''} when they are actually looking for sequence logo visualizations, where each position in the DNA is shown as a stack of nucleotide letters A, T, C, and G, and the size of each letter reflects how often it appears. Without knowledge of the term \textit{``sequence logo,''} they are unable to enter a precise query, as illustrated in the left panel of Figure~\ref{fig:example}.

In contrast to standard multimedia retrieval, visualization retrieval requires a deeper understanding of the data-binding structure within visual representations. The nature of visualization encompasses not just the image, but also domain tasks, user interaction, and human interpretation---which can be subjective and nuanced based on audience. Prior literature on visualization retrieval systems~\cite{chen2024chart2vec, hoque2019searching, li2022structure, oppermann2020vizcommender, setlur2023olio, xiao2023wytiwyr, ye2022visatlas, ying2024vaid, nguyen_geranium} has addressed these challenges by developing specialized similarity measures and query mechanisms, and generally follows a common retrieval loop from the user's perspective: (1) the user articulates a query conveying their intent, (2) the system matches and ranks candidates based on visual or semantic similarity, and (3) the user reviews and refines the search. As this iterative cycle mirrors the sensemaking process~\cite{hunter2025can} and reinforces inquiry-based learning~\cite{lazonder2016meta}, we argue that it is well-suited for educational contexts. Through this process, retrieval supports not only access to examples but also the articulation of intent, a core component of data literacy and critical thinking with data. The next section continues with the implications of visualization retrieval for data literacy in detail.

\section{The Role of Visualization Retrieval in Data Literacy}

Adopting the framework of data literacy as a set of practices across the data lifecycle~\cite{khan2018toward}, we investigate the role and extent of visualization retrieval in data production, consumption, and management. Through this analysis, we highlight the relevance of visualization retrieval as essential literacy infrastructure. In this paper, we use the following working definitions:

\begin{enumerate}
    \item \textbf{Data production}: activities in which learners and practitioners create or modify visualizations to explore data or communicate findings. The learner is a producer or designer of visualizations.
    
    \item \textbf{Data consumption}: activities in which learners and practitioners encounter, interpret, critique, or rely on visualizations created by others. The learner is a reader, interpreter, or critic of visualizations.

    \item \textbf{Data management}: activities in which both learners and educators act as curators of datasets and visualizations, creating and maintaining repositories for future use.
\end{enumerate}

\subsection{Visualization Retrieval and Data Production}

In production-related activities, students are responsible for \textit{authoring} visualizations, including when they start from templates or existing examples. Visualization retrieval can support them by making the design space visible and easy to navigate, providing implementable examples, and expanding their visual vocabulary.

\textbf{Design space exploration}: In the early stage of visualization design, learners must answer questions involving design space such as: \textit{``What visual encodings are suitable for my data (e.g., temporal, multivariate, spatial)?'' } or \textit{``How can I support the analytic task (e.g., comparison, trend detection, anomaly detection)?''} Searching for a certain visualization task or style, the system might return line charts, small multiples, stacked area charts besides the default of bar plots~\cite{xiao2023wytiwyr, ying2024vaid}. In this case, a visualization retrieval system can act as a map of the design space, providing multiple \textit{plausible designs} for the same underlying task. From this exploration, learners can compare trade-offs between options and thus make more intentional design decisions.

\textbf{Inspiration and implementation}: In practice, visualization authoring often follows a find-and-adapt workflow~\cite{interviews, bako_designers_find}. Learners can search for an example that resembles their desired output, then adapt its code or specification to their own data and context. While static galleries like the R Graph Gallery~\cite{RGallery} or Python Graph Gallery~\cite{PyGallery} provide code examples, they rely on manual browsing, forcing users to sift through the collections to find relevant templates. Visualization retrieval systems can significantly streamline this workflow by returning examples directly linked to code, specifications, or notebooks. By allowing learners to filter results by library (e.g., R, Python, React) or data structure, retrieval systems connect abstract design choices to their concrete actualization in code. This not only improves efficiency but reinforces data literacy by helping learners map visual encodings to the data transformations required to produce them.

\textbf{Expanding visual vocabulary}: Retrieving and viewing diverse charts also help expand the \textit{visual vocabulary} for learners. Returning to the earlier genomics example, a learner may type \textit{``ATCG plot''} and, through retrieval, encounter not only sequence logo visualization images, but also domain-specific naming such as \textit{``sequence logo''} and \textit{``sequence motif logo,''} as illustrated in the right panel of Figure~\ref{fig:example}.

\subsection{Visualization Retrieval and Data Consumption}

Data consumption refers to how individuals encounter, interpret, evaluate, and potentially act upon visualizations produced by others. In these activities, learners are not primarily responsible for creating visualizations; instead, they are asked to read, understand, critique, and compare existing visualizations. As understanding visual patterns and pitfalls is an essential component of visualization literacy~\cite{firat2022interactive}, visualization retrieval supports this process by enabling intent-driven exploration of existing visualizations for comparison and critique.

With visualization retrieval, learners can not only explore the design space both with and without purpose~\cite{baigelenov2025visualization}, but also get to articulate their intents through refinements of searches. Systems such as VAID~\cite{ying2024vaid} and Geranium~\cite{nguyen_geranium} support this type of inquiry through multimodal query methods ranging from categorical selection to multimodal inputs (e.g., free text, images, or declarative grammar). Consequently, while static galleries provide exposure to design varieties in a relatively passive manner, visualization retrieval offers a dynamic shift for learners to not only search but articulate their intents and refine the queries over time.

\subsection{Visualization Retrieval and Data Management}

Beyond production and consumption, retrieval aids in data management. As learners build their expertise, they need to organize examples. Retrieval systems can help learners curate personal libraries such as good practices or common pitfalls, and tag examples for future reference; this curation process itself reinforces literacy skills. On the other hand, from an educator's perspective, having a searchable knowledge base of examples helps streamline the development of curriculum materials. Opportunities in this area are presented in the next section.

\section{Opportunities for Research and System Design}

The integration of visualization retrieval into data literacy presents rich opportunities for research and system design. These directions are interlaced between data production, consumption, and management; we propose the following areas to support learners as they move across these practices, addressing both research and education purposes.

\subsection{Integrated Retrieval-Authoring Systems} To fully streamline the process from finding inspiration to exploring the design space and to implementation, we call for development of integrated retrieval-authoring environments. While current workflows often require switching between a search engine (e.g., Google search) and a coding interface, future systems should bridge this gap by:

\textbf{Embedding retrieval in authoring tools and vice-versa:} Developing plugins for environments such as Jupyter or RStudio that retrieve adaptable code templates alongside the user's data, and providing previews on how the resulting visualization would appear. This idea can also be developed as a standalone platform with a large database of examples, where an editor would be built alongside the search functionality.

\textbf{Multimodal search-by-sketch:} Building machine learning models and algorithms that allow learners to query using a rough sketch or an outline of the mental model. This could ease the learning curve for novice learners who lack the vocabulary to query the chart by name but can articulate visually what they want to build.

\subsection{Pedagogical Relevance Modeling}

Defining relevance in an educational context requires moving beyond standard visualization retrieval metrics. Previous user studies indicate that, in addition to high-similarity matches, participants also prioritize the diversity of retrieved results~\cite{nguyen_geranium}. This need for diverse examples is particularly heightened in learning environments. To support this, we need to develop specialized metadata schemas and thesauri that extend current indexing methods and similarity criteria~\cite{nguyen_safire} and capture pedagogical utility.

For example, a schema for pedagogical purpose could encode dimensions of critique and reasoning such as \textit{Cognitive Principles} (e.g., Gestalt proximity) or \textit{Common Pitfalls} (e.g., truncated axes). At the same time, a visualization thesaurus is crucial to bridge the vocabulary gap for learners with different backgrounds and different levels of visualization literacy, by mapping lay descriptions to formal design terminology. For instance, linking a query for ``Manhattan plot'' to ``GWAS significance scatterplot,'' a specialized scatterplot used in genome-wide association studies (GWAS) to visualize the relationship between genomic variants and a specific trait. Establishing this semantic infrastructure is a prerequisite for retrieval systems that can surface not only relevant but also conceptually aligned visualizations.

\subsection{Collaborative Educational Corpora}

Finally, there is an opportunity to build shared, open-access corpora specifically for education. These repositories would link visualizations, datasets, code, and teaching notes, curated by the community to serve as the backend for next-generation educational retrieval systems. Closely related to modeling pedagogical relevance above, capturing these nuanced attributes often requires human expertise that automated extraction tools currently lack. A collaborative corpus would allow educators to contribute these annotations, flagging visualizations that have proven effective in the classroom or that illustrate specific misconceptions. This direction primarily strengthens the management stage of the data lifecycle, while indirectly scaffolding production and consumption by providing a sustainable, well-annotated pool of examples that learners can adapt and critique.

\section{Conclusion}

In this position paper, we have argued that visualization retrieval is a valuable resource for teaching and learning data literacy. Retrieval systems can transform static visualization collections into navigable design spaces that help learners connect informal, task-oriented intents to concrete visual and code implementations. We demonstrated that integrating retrieval into the data lifecycle empowers learners to move beyond passive consumption, enabling them to actively explore design trade-offs, expand their visual vocabulary, and critique representations through comparison. Realizing this potential calls for rethinking how we design and evaluate visualization retrieval systems. Beyond conventional similarity measures, future work should prioritize pedagogical relevance, integrate retrieval more tightly with authoring environments, and invest in community-driven corpora that link visualizations to datasets, code, and teaching goals. Positioning visualization retrieval in this way opens up opportunities to better support inquiry-driven learning, helping learners not only locate relevant visualizations but also articulate their intentions and reason more effectively with data.

\begin{acks}
The authors acknowledge funding by the National Institutes of Health (R01HG011773).
\end{acks}

\bibliographystyle{ACM-Reference-Format}
\bibliography{references}

\end{document}